\documentclass[aps,preprint]{revtex4}
\usepackage{amssymb,epsf}
\usepackage{amsmath}
\usepackage{graphicx}

\begin{document}

\title{Thermodynamics of Taub-NUT/bolt Black Holes in Einstein-Maxwell
Gravity}
\author{M. H. Dehghani$^{1,2}$\footnote{email address:
mhd@shirazu.ac.ir} and A. Khodam-Mohammadi$^{3}$\footnote{email
address: khodam@basu.ac.ir}\footnote{The name of authors has been
written in alphabetic order}}

\affiliation{$^1$Physics Department and Biruni Observatory,
College of Sciences, Shiraz
University, Shiraz 71454, Iran\\
$^2$Research Institute for Astrophysics
and Astronomy of Maragha (RIAAM), Maragha, Iran\\
$^3$Physics Department, College of Sciences, Bu-Ali Sina
University, Hamadan , Iran}

\begin{abstract}
First, we construct the Taub-NUT/bolt solutions of
$(2k+2)$-dimensinal Einstein-Maxwell gravity, when all the factor
spaces of $2k$-dimensional base space $\mathcal{B}$ have positive
curvature. These solutions depend on two extra parameters, other
than the mass and the NUT charge. These are electric charge $q$
and electric potential at infinity $V$. We investigate the
existence of Taub-NUT solutions and find that in addition to the
two conditions of uncharged NUT solutions, there exist two extra
conditions. These two extra conditions come from the regularity of
vector potential at $r=N$ and the fact that the horizon at $r=N$
should be the outer horizon of the NUT charged black hole. We find
that the NUT solutions in $2k+2$ dimensions have no curvature
singularity at $r=N$, when the $2k$-dimensional base space is
chosen to be $\mathbb{CP}^{2k}$. For bolt solutions, there exists
an upper limit for the NUT parameter which decreases as the
potential parameter increases. Second, we study the thermodynamics
of these spacetimes. We compute temperature, entropy, charge,
electric potential, action and mass of the black hole solutions,
and find that these quantities satisfy the first law of
thermodynamics. We perform a stability analysis by computing the
heat capacity, and show that the NUT solutions are not thermally
stable for even $k$'s, while there exists a stable phase for odd
$k$'s, which becomes increasingly narrow with increasing
dimensionality and wide with increasing $V$. We also study the
phase behavior of the 4 and 6 dimensional bolt solutions in
canonical ensemble and find that these solutions have a stable
phase, which becomes smaller as $V$ increases.
\end{abstract}
\maketitle

\section{Introduction}

It has been known for quite long time that the area law of entropy holds for
black holes or black branes in any dimension $2k+2$, when the $U(1)$
isometry group associated with the (Euclidean) timelike Killing vector $%
\partial _{\tau }$ has a fixed point set on the surfaces of $2k$ codimension
\cite{Haw1}. However, in recent years, it has been shown that
entropy can be associated with a more general class of spacetimes
\cite{Haw2,Mann1}. In these spacetimes, the $U(1$) isometry group
can have fixed points on surfaces of any even codimension, and the
spacetime need not be asymptotically flat or asymptotically
anti-de Sitter (AdS). In this more
general class, the entropy is not just a quarter the area of the $2k$%
-dimensional fixed point set. Such situations occur in spacetimes containing
NUT-charges. In $2k+2$ dimensions they not only can have $2k$-dimensional
fixed point sets (called bolts), they also have fixed point sets with
dimension less than $2k$ (called NUT). Here the orbits of the $U(1)$
isometry group develop singularities, which are the gravitational analogues
of Dirac string singularities and are referred to as Misner strings. When
the NUT charge is nonzero, the entropy of a given spacetime includes not
only the entropies of the $2k$-dimensional bolts, but also those of the
Misner strings. In a very basic sense, gravitational entropy can be regarded
as arising from the Gibbs-Duhem relation applied to the path-integral
formulation of quantum gravity \cite{Mann2}. If one calculates the finite
total action $I$ evaluated on the classical solution, then the entropy may
be written as
\begin{equation}
S=\beta (\mathcal{M}-\Gamma _{i}\mathcal{C}_{i})-I  \label{GibD}
\end{equation}
where $\mathcal{C}_{i}$ and $\Gamma _{i}$ are the conserved charges and
their associate chemical potentials respectively.

The original asymptotically locally flat NUT/bolt solutions in
four dimensions have been constructed in Ref. \cite{TNUT}. There
are known extensions of the Taub-NUT/bolt solutions to the case
when a cosmological constant is present. In this case the
asymptotic structure is only locally de Sitter (for positive
cosmological constant) or AdS (for negative cosmological constant)
and the solutions are referred to as Taub-NUT-(A)dS metrics.
Generalizations to higher dimensions follow closely the
four-dimensional case
\cite{Bais,Page,Akbar,Robinson,Awad,Mann3,Astefan}. Taub-NUT
solution of the Einstein equations with multiple NUT parameters
has been investigated in \cite{MS1}. Also, charged Taub-NUT
solution of the Einstein-Maxwell equations in four dimensions is
known \cite{Bril}, and its generalization to six dimensions has
been done in Ref. \cite{Mann4,Awad2}. The existence of NUT charged
solutions of Einstein-Yang-Mills and Einstein-Yang-Mills-Higgs
theory and their thermodynamics have also been considered
\cite{Radu}. Recently, the existence of Taub-NUT/bolt solution in
Gauss-Bonnet and Gauss-Bonnet-Maxwell gravity have been studied by
one of us \cite{DM2,DH}. In this paper we, first, construct the
$(2k+2)$-dimensional NUT/bolt solutions of Einstein gravity in the
presence of electromagnetic field and, second, we calculate their
conserved charges through the use of the counterterm method and
analyze the thermodynamic behavior of these electrically charged
NUT/bolt solutions.

The outline of our paper is as follows. We give a brief review of the
counterterm method in Sec. \ref{Gen-For}. In. Sec. \ref{d-dim} we construct
the $(n+1)$-dimensional Taub-NUT/bolt solutions of the Einstein-Maxwell
gravity. The thermodynamics of electrically charged NUT solutions in $4$%
-dimensions is investigated in Sec. \ref{four-Dim} and that of $6$%
-dimensional solutions is considered in Sec. \ref{six-Dim}. Generalization
of these subjects to the case of $(2k+2)$-dimensional solutions is done in
Sec. \ref{2k+2-therm}. We finish our paper with some concluding remarks.

\section{General Formalism\label{Gen-For}}

The gravitational action for Einstein gravity in $(n+1)$ dimensions in the
presence of cosmological constant and electromagnetic field is
\begin{equation}
I_G=-\frac 1{16\pi }\int_{\mathcal{M}}d^{n+1}x\sqrt{_{-}g}\left( \mathcal{R}%
-2\Lambda -F^{\mu \nu }F_{\mu \nu }\right) -\frac 1{8\pi }\int_{\partial
\mathcal{M}}d^nx\sqrt{_{-}\gamma }K(\gamma ),  \label{Actg}
\end{equation}
where $\Lambda =-n(n-1)/(2l^2)$ is the cosmological constant, $\mathcal{R}$
is the Ricci scalar, $F_{\mu \nu }=\partial _\mu A_\nu -\partial _\nu A_\mu $
is electromagnetic tensor field and $A_\mu $ is the vector potential. The
first term is the Einstein-Hilbert action and the second term is the Gibbons
Hawking boundary term which is chosen such that the variational principle is
well-defined. The manifold $\mathcal{M}$ has metric $g_{\mu \nu }$ and
covariant derivative $\nabla _\mu $. $K$ is the trace of the extrinsic
curvature $K^{\mu \nu }$ of any boundary(ies) $\partial \mathcal{M}$ of the
manifold $\mathcal{M}$, with induced metric(s) $\gamma _{ij}$.

In order to obtain the Einstein-Maxwell equations by the variation of the
volume integral with respect to the fields, one should impose the boundary
condition $\delta A_{\mu }=0$ on $\partial \mathcal{M}$. Thus the action (%
\ref{Actg}) is appropriate to study the grand-canonical ensemble with fixed
electric potential \cite{Cal}. To study the canonical ensemble with fixed
electric charge one should impose the boundary condition $\delta
(n^{a}F_{ab})=0$, and therefore the gravitational action is \cite{Haw3}
\begin{equation}
\overset{\sim }{I}_G=I_G-\frac{1}{4\pi }\int_{\partial
\mathcal{M}_{\infty }}d^{n}x\sqrt{_{-}\gamma }n_{a}F^{ab}A_{b},
\label{Actcan}
\end{equation}
where $n_{a}$\ is the normal to the boundary $\partial \mathcal{M}$. Varying
the action (\ref{Actg}) or (\ref{Actcan}) with respect to the metric tensor $%
g_{\mu \nu }$ and electromagnetic tensor field $F_{\mu \nu }$, with
appropriate boundary condition, the equations of gravitation and
electromagnetic fields are obtained as
\begin{eqnarray}
&& G_{\mu \nu }-\frac{n(n-1)}{2l^{2}}g_{\mu \nu } =8\pi T_{\mu \nu }^{%
\mathrm{(em)}},  \label{gr-field} \\
&& \nabla _{\mu }F^{\mu \nu } =0,  \label{Emeq}
\end{eqnarray}
where $G_{\mu \nu }$ is the Einstein tensor and $T_{\mu \nu }^{\mathrm{(em)}%
}=2F_{\phantom{\lambda}{\mu}}^{\rho }F_{\rho \nu }-\frac{1}{2}F_{\rho \sigma
}F^{\rho \sigma }g_{\mu \nu }$ is the energy-momentum tensor of
electromagnetic field.

In general the action $I_G$ or $\overset{\sim }{I}_G$ are diverged
when evaluated on solutions, as is the Hamiltonian and other
associated conserved charges. One way of eliminating these
divergences is through the use of background subtraction
\cite{BY,BCM}, in which the boundary surface is embedded in
another (background) spacetime, and all quasilocal quantities are
computed with respect to this background, incorporated into the
theory by adding to the action the extrinsic curvature of the
embedded surface. Such a procedure causes the resulting physical
quantities to depend on the choice of reference background;
furthermore, it is not possible in general to embed the boundary
surface into a background spacetime. For asymptotically AdS
solutions, one can instead deal with these divergences via the
counterterm method inspired by AdS/CFT correspondence \cite{Mal}.
This conjecture, which relates the low energy limit of string
theory in asymptotically AdS spacetime and the quantum field
theory on its boundary, has attracted a great deal of attention in
recent years. The equivalence between the two formulations means
that, at least in principle, one can obtain complete information
on one side of the duality by performing computation on the other
side. A dictionary translating between different quantities in the
bulk gravity theory and their counterparts on the boundary has
emerged, including the partition functions of both theories. In
the present context this conjecture furnishes a means for
calculating the action and conserved quantities intrinsically
without reliance on any reference spacetime \cite{Sken,BK,Od2} by
adding additional terms on the boundary that are curvature
invariants of the induced metric. Although there may exist a very
large number of possible invariants one could add in a given
dimension, only a finite number of them are non vanishing as the
boundary is taken to infinity. Its many applications include
computations of conserved quantities for black holes with
rotation, various topologies, rotating black strings with zero
curvature horizons and rotating higher genus black branes \cite
{Deh3}. Although the asymptotic structure of the NUT charged
solutions of Einstein gravity in the presence of a cosmological
constant is only locally (A)dS, the counterterm method has been
employed to calculate the action and the conserved charges of
them. The conserved quantities of asymptotically locally AdS
solutions in four dimensions have been calculated in
\cite{Myers1}, those of higher-dimensional solutions in
\cite{CFM}, and those of asymptotically locally dS solutions in
\cite{MS2}. Also the conserved quantities of the asymptotically
locally AdS solutions have been calculated in \cite{Cai}. Usually,
the counterterm method applies for the case of a specially
infinite boundary, but it was also employed for the computation of
the conserved and thermodynamic quantities in the case of a finite
boundary \cite{DM1}. The general correspondence formula is
\cite{Mal}
\begin{equation}
\int_{\Psi _0}\mathcal{D}\Psi e^{-I_{AdS}[\Psi ]}=\langle \exp \int d^nx%
\mathcal{O}(x)\Psi _0(x)\rangle ,
\end{equation}
where the functional integral on the left hand side is over all the fields $%
\Psi $\ whose asymptotic boundary values are $\Psi _0,$\ and $\mathcal{O}$\
denotes the conformal operators of the boundary conformal field theory. In
the classical limit, the correspondence formula can be written as \cite
{Witten, Gubser}
\begin{equation}
I_{AdS}[\Psi _0]=W_{CFT}[\Psi _0],
\end{equation}
where $I_{AdS}$\ is the classical on-shell action of an AdS field theory,
expressed in terms of the field boundary values $\Psi _0$, and $W_{CFT}$\ is
the CFT effective action. However, one should expect $I_{AdS}$\ to be
divergent as it stands, because of the divergence of the AdS metric on the
AdS horizon. Thus, in order to extract the physically relevant information,
the on-shell action has to be renormalized by adding counterterms, which
cancel the infinities of $I_G$ in the absence of matter. This counterterms
up to nine dimensions is \cite{Kraus}
\begin{eqnarray}
I_{\mathrm{ct}} &=&\frac 1{8\pi }\int_{\partial \mathcal{M}_\infty }d^nx%
\sqrt{-\gamma }\{\frac{n-1}l-\frac{l\Theta (n-3)}{2(n-2)}R  \notag \\
&&\ -\frac{l^3\Theta (n-5)}{2(n-4)(n-2)^2}\left( R_{ab}R^{ab}-\frac
n{4(n-1)}R^2\right) +\frac{l^5\Theta (n-7)}{(n-2)^3(n-4)(n-6)}[\frac{3n+2}{%
4(n-1)}RR_{ab}R^{ab}  \notag \\
&&\ -\frac{n(n+2)}{16(n-1)^2}R^3-2R^{ab}R_{acbd}R^{cd}+\frac{n-2}{2(n-1)}%
R^{ab}\nabla _a\nabla _bR-R^{ab}\square R_{ab}+\frac 1{2(n-1)}R\square R]\}
\label{Ict}
\end{eqnarray}
where $R$, $R_{abcd}$, and $R_{ab}$ are the Ricci scalar, Riemann and Ricci
tensors of the boundary metric $\gamma _{ab}$, and $\Theta (x)$ is the step
function which is equal to one for $x\geq 0$ and zero otherwise. In the
presence of matter, one may encounters with some divergencies. For $n>4$ the
electromagnetic field will cause a power law divergence in the action which
can be removed by a counterterm of the form \cite{Robin}
\begin{eqnarray}
I_{\mathrm{ct}}^{\mathrm{em}}=\frac l{8\pi }\int d^nx\sqrt{_{-}\gamma }
&&\left\{ \frac{\Theta (n-5)}{32}\frac{(n-8)}{(n-4)}F^2+\frac{l^2\Theta (n-7)%
}4[\frac{(5n-11)}{48(n-1)^2(n-2)(n-6)}RF^2\right.  \notag \\
&&\ \ \left. +\frac{(7n-66)}{12(n-6)(n-2)}R_b^aF_{ac}F^{bc}+\frac{(n-8)}{%
12(n-4)^2}(\nabla _aF^{ab})^2\right.  \notag \\
&&\ \ \left. +\frac{(n-12)}{47(n-4)(n-6)}F^{ab}(\nabla _b\nabla
^cF_{ca}-\nabla _a\nabla ^cF_{cb})]\right\}  \label{Iem}
\end{eqnarray}
Thus, the total finite action can be written as a linear combination of the
gravity term (\ref{Actg}) and the counterterms (\ref{Ict}) and (\ref{Iem}).
Having the total finite action, one can use the Brown and York definition of
energy-momentum tensor \cite{BY} to construct a divergence free
stress-energy tensor. This tensor is \textbf{\ }
\begin{eqnarray}
T^{ab} &=&\frac 1{8\pi }\{(K^{ab}-K\gamma ^{ab})-\frac{n-1}l\gamma
^{ab}+\frac l{n-2}(R^{ab}-\frac 12R\gamma ^{ab})  \notag \\
&&\ \ +\frac{l^3\Theta (n-5)}{(n-4)(n-2)^2}[-\frac 12\gamma
^{ab}(R^{cd}R_{cd}-\frac n{4(n-1)}R^2)-\frac n{(2n-2)}RR^{ab}  \notag \\
&&\ \ +2R_{cd}R^{acbd}-\frac{n-2}{2(n-1)}\nabla ^a\nabla ^bR+\nabla
^2R^{ab}-\frac 1{2(n-1)}\gamma ^{ab}\nabla ^2R]+...\}+\frac{\delta I_{%
\mathrm{ct}}^{\mathrm{em}}}{\delta \gamma _{ab}}  \label{Stres}
\end{eqnarray}
The explicit form of the stress-energy tensor due to the
electromagnetic field counterterm will be given in Sec.
\ref{six-Dim}. To compute the conserved charges of the
spacetime, we choose a spacelike surface $\Sigma$ in $\partial \mathcal{%
M}$ with metric $\sigma _{ij}$, and write the boundary metric in
ADM form:
\begin{equation}
\gamma _{ab}dx^{a}dx^{a}=-\mathcal{N}^{2}dt^{2}+\sigma _{ij}\left(
d\varphi ^{i}+\mathcal{V}^{i}dt\right) \left( d\varphi
^{j}+\mathcal{V}^{j}dt\right) ,
\end{equation}
where the coordinates $\varphi ^{i}$ are the angular variables
parameterizing the hypersurface of constant $r$ around the origin,
and $\mathcal{N}$ and $\mathcal{V}^{i}$ are the lapse and shift
functions respectively. The conserved charges associated to a
Killing vector $\xi ^a$ is
\begin{equation}
\mathcal{Q}(\xi )=\int_\Sigma d^{n-1}x\sqrt{\sigma }u^aT_{ab}\xi ^b,
\label{Con}
\end{equation}
where $\sigma $ is the determinant of the metric $\sigma _{ij}$ and $u^a$ is
the normal to the quasilocal boundary hypersurface $\Sigma $.$\ $For
boundaries with timelike Killing vector ($\xi =\partial _t$) one obtains the
conserved mass of the system enclosed by the boundary $\Sigma $. In the
context of AdS/CFT correspondence, the limit in which the boundary $\Sigma $
becomes infinite $(\Sigma _\infty )$ is taken, and the counterterm
prescription ensures that the action and conserved charges are finite. No
embedding of the surface $\Sigma $ in to a reference of spacetime is
required and the quantities which are computed are intrinsic to the
spacetimes.

\section{$(2k+2)$-dimensional Taub-NUT/Bolt Solutions in Einstein-Maxwell
Gravity\label{d-dim}}

The Euclidean section of the ($2k+2$)-dimensional charged Taub-NUT/bolt
spacetime can be written as
\begin{equation}
ds^{2}=F(r)(d\tau +N\mathcal{A})^{2}+F^{-1}(r)dr^{2}+(r^{2}-N^{2})d\Xi _{%
\mathcal{B}},  \label{metr-d}
\end{equation}
where $\tau $ is the coordinate of the fibers $S^{1}$ and $\mathcal{A}$ is
the K\"{a}hler form of the base space $\mathcal{B}$, $N$ is the NUT charge
and $F(r) $ is a function of $r.$ The metric $d\Xi _{\mathcal{B}}$ is a $2k$%
-dimensional base space\ Einstein-K\"{a}hler manifold $\mathcal{B}.$

Here, we consider only the cases where all the factor spaces of $\mathcal{B}$
have positive curvature. Thus, the base space $\mathcal{B}$ may be the
product of $\mathbb{CP}^{k}$ spaces for all values of $k$. For completeness, we give the $1$%
-forms and the metrics of these factor spaces. The $1$-forms and
the metrics of $\mathbb{CP}^{k}$ is
\begin{eqnarray}
\mathcal{A}_{k} &=&2(k+1)\sin ^{2}\xi _{k}(d\psi _{k}+\frac{1}{2k}\mathcal{A}%
_{k-1}),  \label{Ak} \\
d\Sigma _{k}^{2} &=&2(k+1)\left\{ d\xi _{k}^{2}+\sin ^{2}\xi _{k}\cos
^{2}\xi _{k}(d\psi _{k}+\frac{1}{2k}\mathcal{A}_{k-1})^{2}+\frac{1}{2k}\sin
^{2}\xi _{k}d\Sigma _{k-1}^{2}\right\}  \label{CPk}
\end{eqnarray}
where $\mathcal{A}_{k-1}$\ is the K\"{a}hler potential of $%
\mathbb{CP}^{k-1}$. In Eqs. (\ref{Ak}) and (\ref{CPk}) $\xi _{k}$\
and $\psi _{k}$\ are the extra coordinates corresponding to
$\mathbb{CP}^{k}$ with respect to $\mathbb{CP}^{k-1}$. The metric
$\mathbb{CP}^{k}$ is normalized such that, Ricci tensor is equal
to the metric, $R_{\mu \nu }=g_{\mu \nu }$. The $1$-form and the
metric of $\mathbb{CP}^{1}$ ($S^2$) is
\begin{eqnarray}
\mathcal{A}_{1} &=&4\sin ^{2}\xi _{1}d\psi _{1}  \label{A1} \\
d{\Sigma _{1}}^{2} &=&4\left( {d\xi _{1}}^{2}+\sin ^{2}\xi _{1}\cos ^{2}\xi
_{1}{d\psi _{1}}^{2}\right)  \label{CP1}
\end{eqnarray}
and those of $\mathbb{CP}^{k}$ can be constructed through the use of Eqs. (%
\ref{Ak}) and (\ref{CPk}).

The gauge potential has the form \cite{Awad2}
\begin{equation}
A=h(r)(d\tau +N\mathcal{A}),  \label{A}
\end{equation}
where $h(r)$ is a function of $r$. The electromagnetic field equation (\ref
{Emeq}) for the metric (\ref{metr-d}) with vector potential (\ref{A}) is
\begin{equation}
(r^2-N^2)^2h^{\prime \prime }(r)+2kr(r^2-N^2)h^{\prime }(r)-4kN^2h(r)=0
\label{EMd}
\end{equation}
where prime denotes a derivative with respect to $r$. The solution of Eq. (%
\ref{EMd}) may be expressed\textbf{\ }in terms of hypergeometric function $%
_2F_1([a,b],[c],z)$\ in a compact form as
\begin{equation}
h(r)=\frac{qr}{(r^2-N^2)^k}+V(2k-1)N^{2k}\ _2F_1([-\frac 12,-k],[\frac 12],%
\frac{r^2}{N^2})  \label{hrd}
\end{equation}
where $V$\ and $q$ are two arbitrary constants which correspond to electric
potential at infinity and charge respectively. To find the function $F(r)$,
one may use any components of Eq. (\ref{gr-field}). The simplest equation is
the $tt$ component of these equations which can be written as
\begin{equation}
rF^{\prime }(r)+\left( \frac{(2k-1)r^2+N^2}{r^2-N^2}\right) F(r)-\frac{%
l^2+(2k+1)(r^2-N^2)}{l^2}=\frac{(r^2-N^2)}kh^{\prime 2}(r)-\frac{4N^2}{%
(r^2-N^2)}h^2(r),  \label{Eqd}
\end{equation}
with the solution
\begin{eqnarray}
F(r) &=&\frac r{(r^2-N^2)^k}\int^r\frac{(s^2-N^2)^{k-1}}{kl^2s^2}\left\{
l^2(s^2-N^2)^2h^{\prime 2}(s)-4l^2N^2kh^2(s)\right.  \notag \\
&&\left. +k(2k+1)(s^2-N^2)^2+kl^2(s^2-N^2)\right\}
ds-\frac{mr}{(r^2-N^2)^k}, \label{Frd}
\end{eqnarray}
where $h(r)$ is given in Eq. (\ref{hrd}). One may note that the
above solution reduces to the $(2k+2)$-dimensional
solution given in \cite{CFM} in the absence of electromagnetic field ($%
h(r)=0 $).

\subsection{NUT Solutions\label{NUT-k}}

The solutions of Eq. (\ref{Eqd}) describe NUT solutions, if

(I) $F(r_{+}=N)=0.$

(II) $\beta =4\pi /F^{\prime }(r_{+})=4\pi (k+1)N$

(III) $h(r_{+}=N)=0.$

(IV) $F(r)$ should have no positive roots at $r>N$.\newline The
first condition comes from the fact that all the extra dimensions
should collapse to zero at the fixed point set of $\partial
/\partial \tau $, the second one ensures that there is no conical
singularity with a smoothly closed fiber at $r=N$. Of course, if
one use the Misner's argument \cite{Mis}, the period of time
coordinate is found to be different from $4\pi (k+1)N$ when the
space is singular \cite{ Page,Mann4}. Here we consider the period
of Euclidean time coordinate by the elimination of conic
singularities. The third condition comes from the regularity of
vector potential at $r=N$ and the fourth one comes up since $r=N$
should be the outer horizon. Condition (III) gives a relation
between $q$ and $V$ as
\begin{equation}
q_{n}=-\frac{2\sqrt{\pi }N^{2k-1}\Gamma (k+1)}{\Gamma (k-\frac{1}{2})}V_{n}
\label{qn}
\end{equation}
Using the first two conditions with Eq. (\ref{qn}), one finds that
Einstein-Maxwell gravity in even dimensions admits NUT solutions with any
base space when the mass parameter is fixed to be
\begin{equation}
m_{n}=N^{2k-1}\{2B(k)[l^{2}-2(k+1)N^{2}]+4D(k)V_{n}^{2}\}  \label{mn}
\end{equation}
\textbf{\ }where $B_{k}$ and $D(k)$ are \textrm{\ }
\begin{eqnarray}
B(k) &=&\frac{\Gamma (\frac{3}{2}-k)\Gamma (k+1)}{(2k-1)\sqrt{\pi }l^{2}}, \\
D(k) &=&\frac{(-1)^{k-1}\sqrt{\pi }(2k-1)(k-1)\Gamma (k+1)}{k\Gamma (k-\frac{%
1}{2})}.
\end{eqnarray}
Finally, we should apply the fourth condition. As we will see in the next
section, the fourth condition is not necessary in $4$ dimensions. However,
this condition restrict the value of $V_{n}$ to be less than a critical
value $V_{\mathrm{crit}}$ in dimensions higher than four. To find $V_{%
\mathrm{crit}}$ we proceeds as follows. We define the function $g_{\mathrm{%
nut}}(r)$ as the numerator of $F_{\mathrm{nut}%
}(r)=F(q=q_{n},m=m_{n},r)/(r-N) $ which is positive at $r=N$, and solve the
system of two equations
\begin{equation}
\left\{
\begin{array}{l}
g_{\mathrm{nut}}(r)=0 \\
g_{\mathrm{nut}}^{\prime }(r)=0\label{gnut}
\end{array}
\right.
\end{equation}
for the unknown $V$ and $r$. The $V$ obtained by this method is the critical
value $V_{\mathrm{crit}}$. We consider the solutions of system of two
equations (\ref{gnut}) in the following sections for various dimensions.

Now, we calculate the electric charge of the solutions. To determine the
electric field we should consider the projections of the electromagnetic
field tensor on special hypersurfaces. The normal to such hypersurfaces is
\begin{equation}
u^{0}=\frac{1}{\mathcal{N}},\hspace{0.5cm}u^{r}=0,u^{i}=\frac{\mathcal{V}^{i}}{\mathcal{N}},
\end{equation}
where $\mathcal{N}$ and $\mathcal{V}^i$ are the lapse and shift
functions respectively. The electric field is $E^{\mu }=g^{\mu
\rho }F_{\rho \nu }u^{\nu }$, and the electric charge can be found
by calculating the flux of the electric field at infinity,
yielding
\begin{equation}
Q_{k}=(2k-1)(4\pi )^{k-1}q.  \label{Qk}
\end{equation}
The electric potential $\Phi $, measured at infinity with respect to the
horizon, is defined by \cite{Gubser}
\begin{equation}
\Phi =A_{\mu }\chi ^{\mu }\left| _{r\rightarrow \infty }-A_{\mu }\chi ^{\mu
}\right| _{r=r_{+}},  \label{Pot1}
\end{equation}
where $\chi =\partial _{\tau }$ is the null generator of the horizon. We
find
\begin{equation}
\Phi _{n}=(-1)^{k+1}V_{n}.  \label{Phik}
\end{equation}

\subsection{Bolt Solutions}

The conditions for having a regular bolt solution are (I) $F(r=r_b)=0$, (II)$%
\ F^{\prime }(r_b)=[(k+1)N]^{-1}$ and (III) $h(r_b)=0$ with $r_b>N$.
Condition (II) follows from the fact that we want to avoid a conical
singularity at the bolt, together with the fact that the period of $\tau $
will still be $4\pi (k+1)N$. The first and third conditions gives\textbf{\ }
\begin{eqnarray}
m_b &=&\sum_{i=0}^k\binom ki\frac{(-1)^iN^{2i}r_b^{2k-2i-1}}{(2k-2i-1)}+%
\frac{(2k+1)}{l^2}\sum_{i=0}^{k+1}\binom{k+1}i\frac{(-1)^iN^{2i}r_b^{2k-2i+1}%
}{(2k-2i+1)}  \notag \\
&&\ +\int^{r_b}\frac{(r^2-N^2)^{k-1}}{kr^2}[(r^2-N^2)^2h^{\prime
2}(r)-4N^2kh^2(r)]dr.  \label{mB-d}
\end{eqnarray}
where $V_b$ is the solution of $h(r_b)=0$. The second condition together
with Eq. (\ref{mB-d}) gives the following equation for $r_b$
\begin{equation}
kl^2r_b^3-k(k+1)N[(2k+1)(r_b^2-N^2)+l^2]r_b^2-(k+1)(2k-1)^2l^2V_b^2N(r_b^2-N^2)=0.
\label{rB-k}
\end{equation}
Equation (\ref{rB-k}) for fixed values of $l$ and $V_b$ is a quartic
equation in $r_b$ and has two real solutions greater than $N$ provided $%
N<N_{\max }$ and one real solution greater than $r_b$ if
$N=N_{\max }$. Numerical calculation shows that $N_{\max }$
decreases as $V$ increases in various dimensions. We will give the
equation which is satisfied by $N_{\max }$ in four and six
dimensions in Secs. \ref{four-Dim} and \ref{six-Dim} respectively.
The electric charge is the same as the NUT solutions, Eq.
(\ref{Qk}), while the electric potential may be obtained by use of
Eq. (\ref{Pot1}) as
\begin{equation}
\Phi _b=-\frac{(-1)^kqr_b}{(2k-1)N_{\quad \quad 2}^{2k}F_1([-\frac
12,-k],[\frac 12],\frac{r_b^2}{N^2})}.  \label{PhiB-k}
\end{equation}

\section{Thermodynamics of 4-dimensional Solutions\label{four-Dim}}

In four dimensions, the functions $h(r)$ and $F(r)$ given in Eqs. (\ref{hrd}%
) and (\ref{Frd}) are
\begin{eqnarray}
h(r) &=&\frac{qr}{r^{2}-N^{2}}+V\frac{r^{2}+N^{2}}{r^{2}-N^{2}},  \label{hr4}
\\
F(r) &=&\frac{%
r^{4}+(l^{2}-6N^{2})r^{2}-ml^{2}r+(4N^{2}V^{2}+N^{2}-q^{2})l^{2}-3N^{4}}{%
l^2(r^{2}-N^{2})}
\end{eqnarray}

\subsection{NUT Solutions}

Since $m_{n}$ and $q_{n}$ of Eqs. (\ref{mn}) and (\ref{qn}) become
\begin{equation}
m_{n}=2Nl^{-2}(l^{2}-4N^{2}),
\end{equation}
\begin{equation}
q_{n}=-2NV_{n},
\end{equation}
the function $F(r)$ for NUT solution can be written as
\begin{equation*}
F_{n}(r)=\frac{r^{2}(r+N)+l^{2}(r-N)-N^{2}(5r-3N)}{(r+N)l^{2}}.
\end{equation*}
It is notable that $F_{n}(r)$ is independent of $q$ and $V$, and therefore
there is no restriction on $V_{n}$ from the fourth condition. The electric
charge and potential are
\begin{eqnarray}
Q &=&q, \\
\Phi _{n} &=&-\frac{q}{2N}
\end{eqnarray}

Using Eqs. (\ref{Actg}) and (\ref{Ict}), the Euclidean actions in the
grand-canonical and canonical ensembles can be calculated as
\begin{eqnarray}
I_{n} &=&\frac{4\pi N^{2}}{l^{2}}(l^{2}-2N^{2}+2l^{2}V^{2}), \\
\tilde{I}_{n} &=&\frac{4\pi N^{2}}{l^{2}}(l^{2}-2N^{2}-2l^{2}V^{2})
\end{eqnarray}
Also the total mass may be calculated as
\begin{equation}
\mathcal{M}_n=\frac{m_{n}}{2},
\end{equation}
and as in the case of uncharged solutions \cite{CFM} the total
angular momentum is zero. The Gibbs and Helmholz free energies are
\begin{eqnarray}
G(T,\Phi ) &=&\frac{I}{\beta }=\frac{32(2\Phi ^{2}+1)l^{2}\pi ^{2}T^{2}-1}{%
512l^{2}\pi ^{3}T^{3}}, \\
F(T,Q) &=&\frac{\tilde{I}}{\beta }=-\frac{1}{512}\frac{1-32l^{2}\pi
^{2}T^{2}+1024Q^{2}l^{2}\pi ^{4}T^{4}}{l^{2}\pi ^{3}T^{3}},
\end{eqnarray}
where $T$ is the Hawking temperature which is inverse of $\beta _{n}=8\pi N$%
. The entropy can be obtained as
\begin{equation}
S_{n}=-\left( \frac{\partial G}{\partial T}\right) _{\Phi }=-\left( \frac{%
\partial F(T,Q)}{\partial T}\right) _{Q}=\frac{4\pi N^{2}}{l^{2}}%
(l^{2}-6N^{2}+2l^{2}V^{2}) .  \label{S-4}
\end{equation}
The entropy can also be obtained through the use of Gibbs-Duhem relation (%
\ref{GibD}), where $\mathcal{C}_{i}=Q$ and $\Gamma _{i}=\Phi $. As one can
see from Eq. (\ref{S-4}), the entropy is positive for 1) $l^{2}>6N^{2}$ and
also 2) $l^{2}<6N^{2}$ provided $\left| V_{n}\right| >(2l)^{-1}\sqrt{%
12N^{2}-2l^{2}}$. It is worth to mention that the second case occurs only in
the case of 4-dimensional NUT solution in the presence of electromagnetic
field.

The Smarr-type formula for mass versus extensive quantities $S$
and$\,Q$ may be written as
\begin{equation}
\mathcal{M}_n=\frac{\sqrt{Z}(-4Z+l^2)}{l^2},
\end{equation}
where $Z$\ is
\begin{equation}
Z=\frac l{12}(1\pm \sqrt{l^2+12Q^2-6S/\pi })  \label{Z4}
\end{equation}
One may then regard the parameters $S$ and $Q$ as a set of extensive
parameters for the mass $\mathcal{M}(S,Q)$ and define the intensive
parameters conjugate to $S$ and $Q$, which are the temperature and the
electric potential respectively. One obtains
\begin{eqnarray}
T_n &=&\left( \frac{\partial \mathcal{M}_n}{\partial S}\right) _Q=\frac
1{8\pi N},  \nonumber \\
\ \ \Phi _n &=&\left( \frac{\partial M}{\partial Q}\right)
_S=-\frac q{2N} \label{TP4}
\end{eqnarray}
for $+$ and $-$ sign in Eq. (\ref{Z4}) provided $12N^2>l^2$ \ and
$12N^2<l^2$ respectively. Equations (\ref{TP4}) show that the the
quantities $T_n$ and $\Phi_n$ coincide with those which was
calculated in Sec. (\ref{d-dim}). Thus the thermodynamic
quantities calculated in Sec. (\ref{d-dim}) for Nut solution,
satisfy the first law of thermodynamics $dM=TdS+\Phi dQ$.

The stability of a thermodynamic system with respect to the small variations
of the thermodynamic coordinates, is usually performed by analyzing the
behavior of the entropy near equilibrium. The local stability in any
ensemble requires that $S(\mathbf{Y})$ be a concave function of its
extensive variables or that its Legendre transformation is a convex function
of the intensive variables\textbf{.} The stability can also be studied by
the behavior of the energy $M(\mathbf{X})$ which should be a convex function
of its extensive variable. In our case the mass $M(S,Q)$ is a function of
entropy and charge. The number of thermodynamic variables depends on the
ensemble that is used. In the canonical ensemble, the charge is a fixed
parameter, and therefore the positivity of the heat capacity $%
C_{Q}=T(\partial S/\partial T)_{Q}$ is sufficient to ensure local stability.
The heat capacity $C_{Q}$ at constant charge is

\begin{equation}
C_{n}=\frac{8\pi N^{2}}{l^{2}}(12N^{2}-l^{2})
\end{equation}
which is positive provided $l^{2}<12N^{2}$. Thus, the NUT solution is stable
in two cases. The first case is when $6N^{2}<l^{2}<12N^{2}$ for which both
the entropy and the heat capacity are positive. The second case is when $%
l^{2}<6N^{2}$ provided $\left| V_{n}\right| >(2l)^{-1}\sqrt{12N^{2}-2l^{2}}$%
. This second case is a stable phase for charged NUT solutions ,
which does not occur in the absence of electromagnetic field. That
is, by turning on the electromagnetic field, the stable phase of
the four-dimensional NUT solution gets a new zone.

\subsection{Bolt Solutions}

In 4 dimensions, the bolt mass, charge and electric potential may be written
from Eqs. (\ref{mB-d}), (\ref{hr4}), and (\ref{PhiB-k}) as
\begin{eqnarray}
m_{b} &=&\frac{r_{b}^{4}+(l^{2}-6N^{2})r_{b}^{2}+N^{2}(l^{2}-3N^{2})}{%
l^{2}r_{b}}-\frac{V^{2}(r_{b}^{2}-N^{2})^{2}}{r_{b}^{3}}, \\
q_{b} &=&-\frac{V(r_{b}^{2}+N^{2})}{r_{b}}, \\
\Phi _{b} &=&-\frac{qr_{b}}{r_{b}^{2}+N^{2}}.
\end{eqnarray}
where $r_{b}$ is the solution of the following equation
\begin{equation}
6Nr_{b}^{4}-l^{2}r_{b}^{3}+2N(l^{2}-3N^{2}+l^{2}V^{2})r_{b}^{2}-2l^{2}V^{2}N^{3}=0.
\label{rb4}
\end{equation}
For fixed values of $l$ and $V$, Eq. (\ref{rb4}) has two real solutions
greater than $N$ provided $N<N_{\max }$ and one real solution $r_{b}>N$ \
for $N=N_{\max }$ where $N_{\max }$ is the smallest root of the following
equation
\begin{eqnarray}
&&62208N^{10}-82944(1-V^{2})l^{2}N^{8}+(41472V^{4}-27648V^{2}+41904)l^{4}N^{6}
\notag \\
&&+(9216V^{6}+9216V^{4}-4464V^{2}-9648)l^{6}N^{4}+(768V^{8}+3072V^{6}  \notag
\\
&&+3024V^{4}+1632V^{2}+912)l^{8}N^{2}-(16V^{6}+48V^{4}+21V^{2}+16)l^{10}=0
\label{eqNmax4}
\end{eqnarray}
Thus, we have bolt solution for $N\leq N_{\max }$. For example,
for $l=1$ and $V=1$, the maximum value of $N_{\max }$ is $0.1033$.
It is a matter of numerical calculation to show that $N_{\max }$\
decreases as $V$ increases. The finite actions in grand-canonical
and canonical ensembles are

\begin{eqnarray}
I_{b} &=&-\pi \frac{\lbrack
r_{b}^{4}-l^{2}r_{b}^{2}+N^{2}(3N^{2}-l^{2})]r_{b}^{2}-(r_{b}^{4}+4N^{2}r_{b}^{2}-N^{4})l^{2}V^{2}%
}{(3r_{b}^{2}-3N^{2}+l^{2})r_{b}^{2}+(r_{b}^{2}-N^{2})l^{2}V^{2}}, \\
\tilde{I}_{b} &=&-\pi \frac{\lbrack
r_{b}^{4}-l^{2}r_{b}^{2}+N^{2}(3N^{2}-l^{2})]r_{b}^{2}+(3r_{b}^{4}+N^{4})l^{2}V^{2}%
}{(3r_{b}^{2}-3N^{2}+l^{2})r_{b}^{2}+(r_{b}^{2}-N^{2})l^{2}V^{2}},
\end{eqnarray}
while the entropy and specific heat are calculated as
\begin{eqnarray}
S_{b} &=&\pi \frac{3r_{b}^{4}+r_{b}^{2}(l^{2}-12N^{2})+N^{2}(l^{2}-3N^{2})}{%
(3r_{b}^{2}-3N^{2}+l^{2})r_{b}^{2}+(r_{b}^{2}-N^{2})l^{2}V^{2}}r_{b}^{2}
\notag \\
&&\ +\pi \frac{\lbrack r_{b}^{4}+4N^{2}r_{b}^{2}-N^{4}]l^{2}V^{2}}{%
(3r_{b}^{2}-3N^{2}+l^{2})r_{b}^{2}+(r_{b}^{2}-N^{2})l^{2}V^{2}},
\end{eqnarray}
\begin{eqnarray*}
C_{b} &=&\ \left\{
[4N(r_{b}^{2}+N^{2})l^{2}V^{2}+36Nr_{b}^{4}-5l^{2}r_{b}^{3}\
+4Nr_{b}^{2}(2l^{2}-3N^{2})-l^{2}N^{2}r_{b}]l^{2}r_{b}^{5}\right\}
^{-1}\times \\
&&2N\pi \lbrack -\left(
48N^{3}r_{b}^{6}l^{4}+6Nr_{b}^{8}l^{4}+6N^{9}l^{4}+48N^{7}r_{b}^{2}l^{4}-108N^{5}r_{b}^{4}l^{4}\right) V^{4}-(264N^{3}r_{b}^{8}l^{2}
\\
&&+36Nr_{b}^{10}l^{2}+12Nr_{b}^{8}l^{4}+216N^{7}l^{2}r_{b}^{4}-45N^{2}r_{b}^{7}l^{4}+60N^{9}l^{2}r_{b}^{2}-76N^{5}r_{b}^{4}l^{4}-5r_{b}^{9}l^{4}
\\
&&+5N^{6}r_{b}^{3}l^{4}-16N^{7}r_{b}^{2}l^{4}-960N^{5}l^{2}r_{b}^{6}+45N^{4}r_{b}^{5}l^{4}+112N^{3}r_{b}^{6}l^{4})V^{2}+264N^{3}r_{b}^{8}l^{2}
\\
&&+15r_{b}^{11}l^{2}+360N^{7}r_{b}^{6}-213N^{2}r_{b}^{9}l^{2}-36Nr_{b}^{10}l^{2}-396N^{5}r_{b}^{8}-6Nr_{b}^{8}l^{4}+72N^{7}l^{2}r_{b}^{4}
\\
&&-12N^{5}l^{2}r_{b}^{6}-10N^{5}r_{b}^{4}l^{4}+5r_{b}^{9}l^{4}+7N^{4}r_{b}^{5}l^{4}-63N^{4}l^{2}r_{b}^{7}+1368N^{3}r_{b}^{10}-54Nr_{b}^{12}
\\
&&-126N^{9}r_{b}^{4}-16N^{3}r_{b}^{6}l^{4}+12N^{2}r_{b}^{7}l^{4}-27N^{6}l^{2}r_{b}^{5}].
\end{eqnarray*}
\begin{figure}[tbp]
\epsfxsize=6cm \centerline{\epsffile{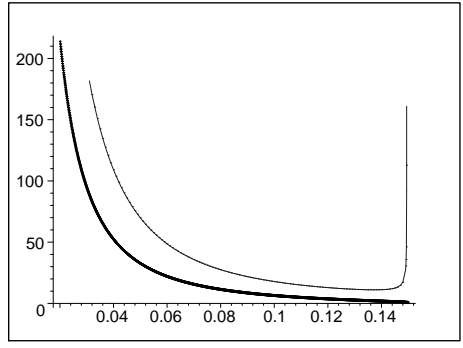}} \caption{Entropy
(bold line) and heat capacity (solid line) of 4-dimensional bolt
solution for $l=1$ and $V=0$ in the allowed range of $N$.}
\label{CS4a}
\end{figure}
\begin{figure}[tbp]
\epsfxsize=6cm \centerline{\epsffile{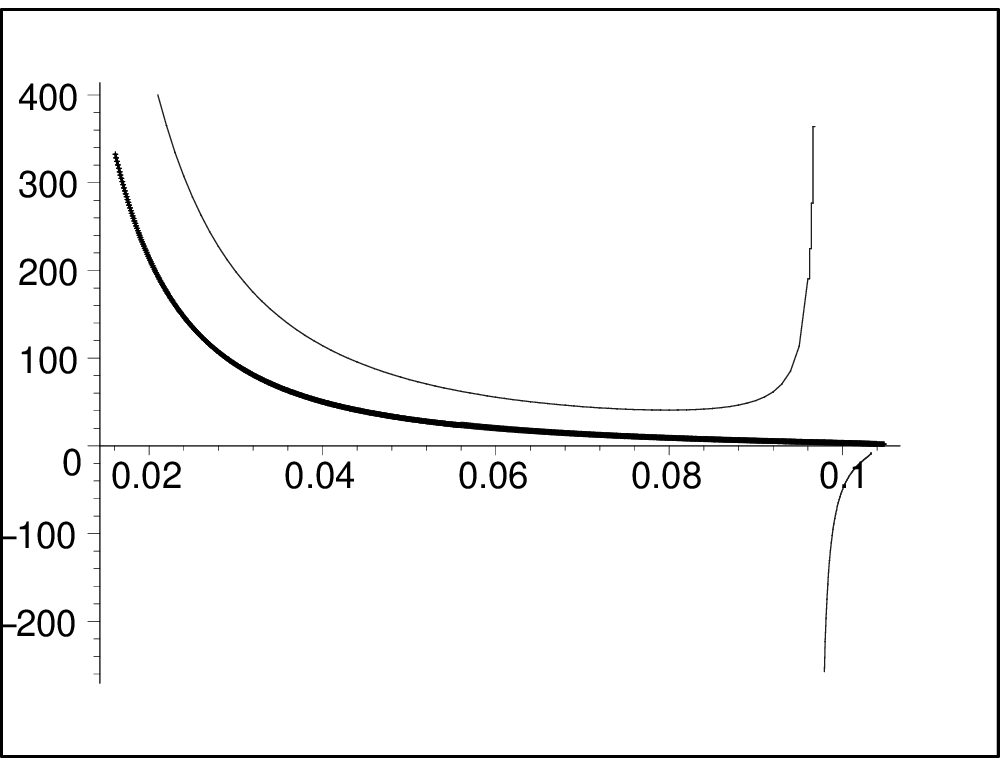}} \caption{Entropy
(bold line) and heat capacity (solid line) of 4-dimensional bolt
solution for $l=1$ and $V=1$ in the allowed range of $N$.}
\label{CS4b}
\end{figure}
\begin{figure}[tbp]
\epsfxsize=6cm \centerline{\epsffile{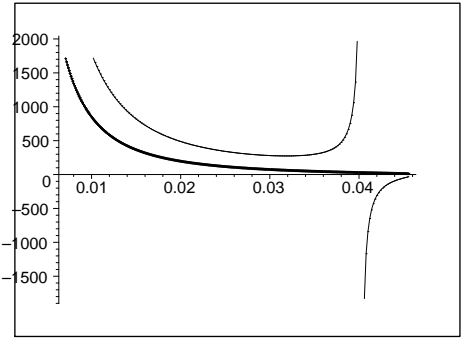}} \caption{Entropy
(bold line) and heat capacity (solid line) of 4-dimensional bolt
solution for $l=1$ and $V=3$ in the allowed range of $N$.}
\label{CS4c}
\end{figure}
In Figs. \ref{CS4a}, \ref{CS4b} and \ref{CS4c}, the entropy and specific
heat are plotted for $l=1$, $V=0$, $V=1$ and $V=3$ up to $N_{\max }$ for
outer horizons of the bolt solutions respectively. As one can see from these
figures, the bolt solution is stable in the range $0<N<$ $N_{\mathrm{st}%
}<N_{\max }$, while $N_{\mathrm{st}}$ is less than $N_{\max }$ for $V\neq 0$%
, and equal to $N_{\max }$ for $V=0$. That is, the uncharged solution is
stable in the whole allowed range $0<N<$ $N_{\max }$, while the stable phase
becomes smaller as $V$ increases.

\section{Thermodynamics of 6-dimensional Solutions\label{six-Dim}}

In six dimensions the base space $\mathcal{B}$ can be the $4$-dimensional
space $\mathbb{CP}^2$\ or the product of two $2$-dimensional spaces $%
S^2\times S^2$. From Eqs. (\ref{hrd}) and (\ref{Frd}) the functions $h(r)$
and $F(r)$ may be written as
\begin{equation}
h(r)=\frac{qr}{(r^2-N^2)^2}-\frac{V(r^4-6r^2N^2-3N^4)}{(r^2-N^2)^2}.
\end{equation}

\begin{eqnarray}
F(r) &=&\frac{3r^6+(l^2-15N^2)r^4+3N^2(15N^2-2l^2)r^2+3N^4(5N^2-l^2)}{%
l^2(r^2-N^2)^2}-\frac{mr}{(r^2-N^2)^2}  \notag \\
&&-\frac{3r^2-N^2}{2(r^2-N^2)^4}q^2-4N^2\frac{r^6+15N^2r^4-9N^4(r^2-N^2)}{%
(r^2-N^2)^4}V^2-\frac{16N^2r^3}{(r^2-N^2)^4}qV.  \label{F6}
\end{eqnarray}
\subsection{NUT Solutions}
\begin{figure}[tbp]
\epsfxsize=7cm \centerline{\epsffile{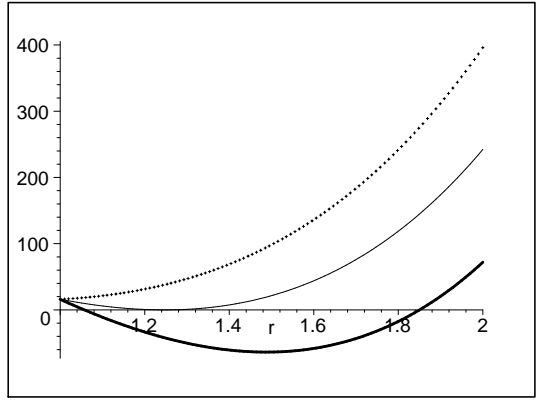}}
\caption{$F_n(r)$ versus $r$ for $V=3<V_{\mathrm{crit}}$ (point), $V=V_{%
\mathrm{crit}}=4.67$ (line) and $V=6>V_{\mathrm{crit}}$
(bold-line).} \label{Fig6d}
\end{figure}

The mass and charge parameters of the NUT solutions may be
obtained from Eqs. (\ref{mn}) and (\ref{qn}) as
\begin{eqnarray}
m_n &=&-\frac{8N^3}{3l^2}(l^2-6N^2+9l^2V_n^2), \\
q_n &=&8N^3V_n.
\end{eqnarray}
Substituting the above parameters in Eq. (\ref{F6}) one obtains
\begin{eqnarray}
F_n(r) &=&\frac{(r-N)}{3(r+N)^4l^2}[-12N^2V_n^2l^2(r-N)+r^3(3r^2+24N^2+l^2)
\notag \\
&&\ \ \ +5Nr^2(3r^2+l^2)+rN^2(-27N^2+7l^2)+3N^3(-5N^2+l^2)].
\end{eqnarray}
It is worth to note that the Taub-NUT solution has not curvature singularity
at $r=N$ for $\mathcal{B}=\mathbb{CP}^2$, while for $\mathcal{B}=S^2\times
S^2$ the metric\ has curvature singularity. Since $F_n(r)$ depends on $V_n$,
the fourth condition for NUT solution restricts the allowed values of $V_n$.
As we discussed in Sec. \ref{d-dim}, one may obtained the critical value of $%
V_{\mathrm{crit}}$ by solving the system of two equations (\ref{gnut}). To
be more clear, we obtain $V_{\mathrm{crit}}$ for asymptotically flat ($%
l\rightarrow \infty $) NUT solutions. The system of two equations (\ref{gnut}%
) becomes
\begin{equation*}
\left\{
\begin{array}{l}
(r+3N)(r+N)^2-12N^2(r-N)V_{\mathrm{crit}}^2=0, \\
(3r+7N)(r+N)-12N^2V_{\mathrm{crit}}^2=0
\end{array}
\right.
\end{equation*}
with the following solution for $V_{\mathrm{crit}}$
\begin{equation}
V_{\mathrm{crit}}=\frac{\sqrt{66+30\sqrt{5}}}6\approx 1.92.
\end{equation}
For arbitrary values of $l$, one may find the critical value of $V$
numerically. For $l=1$ and $N=1$ the critical value of potential which is
obtained by the above method is $V_{\mathrm{crit}}=4.67$. This can be seen
in Fig. \ref{Fig6d} which shows the function $F_n(r)$ as a function of $r$
for various values of $V$ including $V=V_{\mathrm{crit}}$.
The electric charge and potential are
\begin{eqnarray}
Q_n &=&12\pi q, \\
\Phi_n &=&\frac q{8N^3}.
\end{eqnarray}
The Euclidean actions can be calculated by use of  Eqs.
(\ref{Actg}), (\ref{Ict}) and (\ref{Iem})  as
\begin{equation}
I_n=\frac{8\pi N^3\beta }{3l^2}(l^2-4N^2-18l^2V^2),
\end{equation}
It is notable to mention that the electromagnetic field will cause
a power law divergence $16\pi V^{2}N^{2}r$ in action which is
removed by the counterterm (\ref{Iem}). The stress-energy tensor
corresponding to the counterterm (\ref{Iem}) in six dimension is
\begin{equation}
T_{a b}^{\mathrm{(em)}%
}=\frac{1}{4\pi}(F_{\phantom{\lambda}{a}}^{c }F_{c b
}-\frac{1}{4}F_{c d }F^{c d }g_{a b})
\end{equation}
The mass due to this counterterm is
\begin{equation}
\mathcal{M}^{\mathrm{(em)}}_{\mathrm{ct}}=16\pi N^2 V^2 r,
\end{equation}
which removes the power law divergence of the mass due to
electromagnetic field. The total mass is
\begin{equation}
\mathcal{M}_n=4\pi m_n
\end{equation}
The Smarr-type formula for mass is
\begin{equation}
\mathcal{M}_n=\frac{6144\pi ^2Z^4-1024\pi ^2l^2Z^3-l^2Q^2}{96\pi Z^{3/2}l^2},
\end{equation}
where $Z=N^2$ is a function of $Q$ and $S$ which may be obtained by the
following equation
\begin{equation}
640\pi ^2Z^4-96\pi ^2l^2Z^3-Sl^2Z-\frac 3{16}l^2Q^2=0.
\end{equation}
The entropy and specific heat are
\begin{eqnarray}
S_n &=&\frac{32\pi ^2N^4}{l^2}(20N^2-3l^2-54l^2V^2),  \label{S6} \\
C_n &=&\frac{384\pi ^2N^4}{l^2}(l^2-10N^2-9l^2V^2).  \label{C6}
\end{eqnarray}
As one can see from Eq. \ref{S6}, for $\left| V\right| <(6l)^{-1}\sqrt{%
(40/3)N^2-2l^2}$\ and $\left| N\right| \geq \sqrt{0.15}l$, the entropy is
positive,\textrm{\ }but $C_n$ is negative in this range. Thus, $S_n$ and $%
C_n $ are not positive simultaneously, and the NUT black hole is completely
unstable in this ensemble as in the case of uncharged solution \cite{CFM}.

\subsection{Bolt Solutions}

The parameters $m_{b},$ $q_{b}$ and $\beta _{b}$ are
\begin{eqnarray}
m_{b} &=&\frac{1}{3l^{2}r_{b}}%
[3r_{b}^{6}+(l^{2}-15N^{2})r_{b}^{4}-3N^{2}(2l^{2}-15N^{2})r_{b}^{2}-3N^{4}(l^{2}-5N^{2})
\\
&&-\frac{3V^{2}}{2r_{b}^{3}}%
(r_{b}^{6}+3N^{2}r_{b}^{4}+15N^{4}r_{b}^{2}-3N^{6}), \\
q_{b} &=&\frac{V(r_{b}^{4}-6N^{2}r_{b}^{2}-3N^{4})}{r_{b}}, \\
\beta _{b} &=&\frac{8\pi l^{2}r_{b}^{3}}{%
2(5r_{b}^{2}-5N^{2}+l^{2})r_{b}^{2}+9(r_{b}^{2}-N^{2})l^{2}V^{2}}.
\end{eqnarray}
where $r_{b}$ is real root of the following equation
\begin{equation}
30Nr_{b}^{4}-2l^{2}r_{b}^{3}+3(2l^{2}-10N^{2}+9l^{2}V^{2})Nr_{b}^{2}-27N^{3}l^{2}V^{2}=0.
\label{rb6}
\end{equation}
Equation (\ref{rb4}) has two real solutions greater than $N$ provided $%
N<N_{\max }$ and one real solution $r_{b}>N$ \ for $N=N_{\max }$ where $%
N_{\max }$ is the smallest root of the following equation
\begin{eqnarray}
&&9\times (10)^{5}N^{10}+(10)^{4}\times
(324V^{2}-72)l^{2}N^{8}+(10)^{3}\times (4374V^{4}-648V^{2}+217)l^{4}N^{6}
\notag \\
&&+(10)^{2}\times (26244V^{6}+5832V^{4}-999V^{2}-294)l^{6}N^{4}+(10)\times
(59049V^{8}+52488V^{6}  \notag \\
&&+14823V^{4}+2052V^{2}+156)l^{8}N^{2}-(729V^{6}+486V^{4}+81V^{2}+8)l^{10}=0.
\label{eqNmax6}
\end{eqnarray}
As in the case of 4-dimensional solutions, $N_{\max }$\ decreases as $V$
increases.

The electric potential is
\begin{equation}
\Phi _{b}=-\frac{qr_{b}}{r_{b}^{4}-6N^{2}r_{b}^{2}-3N^{4}}.
\end{equation}
The finite action, entropy and heat capacity can be obtained as
\begin{eqnarray}
I_{b} &=&\frac{8\pi ^{2}}{3}\frac{%
r_{b}^{2}[-3r_{b}^{6}+(l^{2}+5N^{2})r_{b}^{4}+3N^{2}(5N^{2}-2l^{2})r_{b}^{2}+3N^{4}(5N^{2}-l^{2})]%
}{2(5r_{b}^{2}-5N^{2}+l^{2})r_{b}^{2}+9(r_{b}^{2}-N^{2})l^{2}V^{2}}  \notag
\\
&&+12\pi ^{2}l^{2}V^{2}\frac{%
(r_{b}^{6}-15N^{2}r_{b}^{4}-21N^{4}r_{b}^{2}+3N^{6})}{%
2(5r_{b}^{2}-5N^{2}+l^{2})r_{b}^{2}+9(r_{b}^{2}-N^{2})l^{2}V^{2}},
\end{eqnarray}

\begin{eqnarray}
S_b &=&-\frac{8\pi ^2}3\frac{%
[15r_b^6+(3l^2-65N^2)r_b^4+3N^2(55N^2-6l^2)r_b^2+9N^4(5N^2-l^2)]r_b^2}{%
2(5r_b^2-5N^2+l^2)r_b^2+9(r_b^2-N^2)l^2V^2}  \notag \\
&&\ \ \ \ +36\pi ^2l^2V^2\frac{r_b^6-15r_b^4N^2-21r_b^2N^4+3N^6}{%
2r_b^2(5r_b^2-5N^2+l^2)+9(r_b^2-N^2)l^2V^2},  \label{Sb6}
\end{eqnarray}
\begin{eqnarray}
C_b &=&3\pi ^2N\{l^2r_b^5[\left( 27l^2N^5+54N^3r_b^2l^2-9Nl^2r_b^4\right)
V^2-90N^5r_b^2+24N^3r_b^2l^2-r_bl^2N^4-8Nl^2r_b^4  \notag \\
&&\
+220N^3r_b^4-10l^2r_b^3N^2-50Nr_b^6+3r_b^5l^2]%
\}^{-1}[-3600N^{11}l^2r_b^4-440N^4r_b^9l^4+9800N^6l^2r_b^9  \notag \\
&&\
+900N^{10}r_b^5l^2+20Nr_b^{12}l^4-156N^8r_b^5l^4-14880N^7l^2r_b^8-288N^3r_b^{10}l^4+18000N^5r_b^{10}l^2
\notag \\
&&\
+1152N^7r_b^6l^4-1680N^3l^2r_b^{12}-4600N^4l^2r_b^{11}+160N^2r_b^{11}l^4-8280N^9r_b^6l^2-576N^6r_b^7l^4
\notag \\
&&\ +324N^9l^4r_b^4+840N^5{r}%
^8l^4+4020N^8r_b^7l^2+200Nl^2r_b^{14}+180N^2r_b^{13}l^2-60r_b^{15}l^2-12r_b^{13}l^4
\notag \\
&&\
+23100N^9r_b^8+500Nr_b^{16}-204800N^7r_b^{10}+62500N^5r_b^{12}+400N^3r_b^{14}+9900N^{13}r_b^4
\notag \\
&&\
+6000N^{11}r_b^6+(31536N^7r_b^6l^4-20412N^9l^4r_b^4+7938N^8r_b^5l^4+900Nl^2r_b^{14}-6156N^6r_b^7l^4
\notag \\
&&\
-12312N^5r_b^8l^4-6264N^3r_b^{10}l^4+378N^{10}r_b^3l^4+5364N^4r_b^9l^4+180Nr_b^{12}l^4-1944N^{11}r_b^2l^4
\notag \\
&&\ +11340N^{13}r_b^2l^2-386100N^9r_b^6l^2+1746N^2r_b^{11}l^4+159840N^7l^2{r}%
^8+116640N^{11}l^2r_b^4  \notag \\
&&\
-52380N^5r_b^{10}l^2-34560N^3l^2r_b^{12}-54r_b^{13}l^4)V^2+(19440N^7r_b^6l^4-10692N^3r_b^{10}l^4
\notag \\
&&\ -14823N^5r_b^8l^4-117369N^9l^4{r}%
^4+37908N^{11}r_b^2l^4+405Nr_b^{12}l^4+2187N^{13}l^4)V^4].
\end{eqnarray}

\begin{figure}[tbp]
\epsfxsize=6cm \centerline{\epsffile{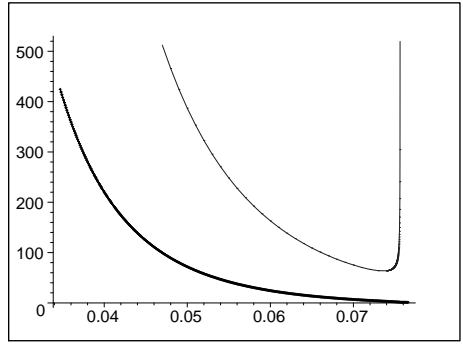}} \caption{Entropy
(bold line) and heat capacity (solid line) of 6-dimensional bolt
solution for $l=1$ and $V=0$ in the allowed range of $N$.}
\label{CS6a}
\end{figure}
\begin{figure}[tbp]
\epsfxsize=6cm \centerline{\epsffile{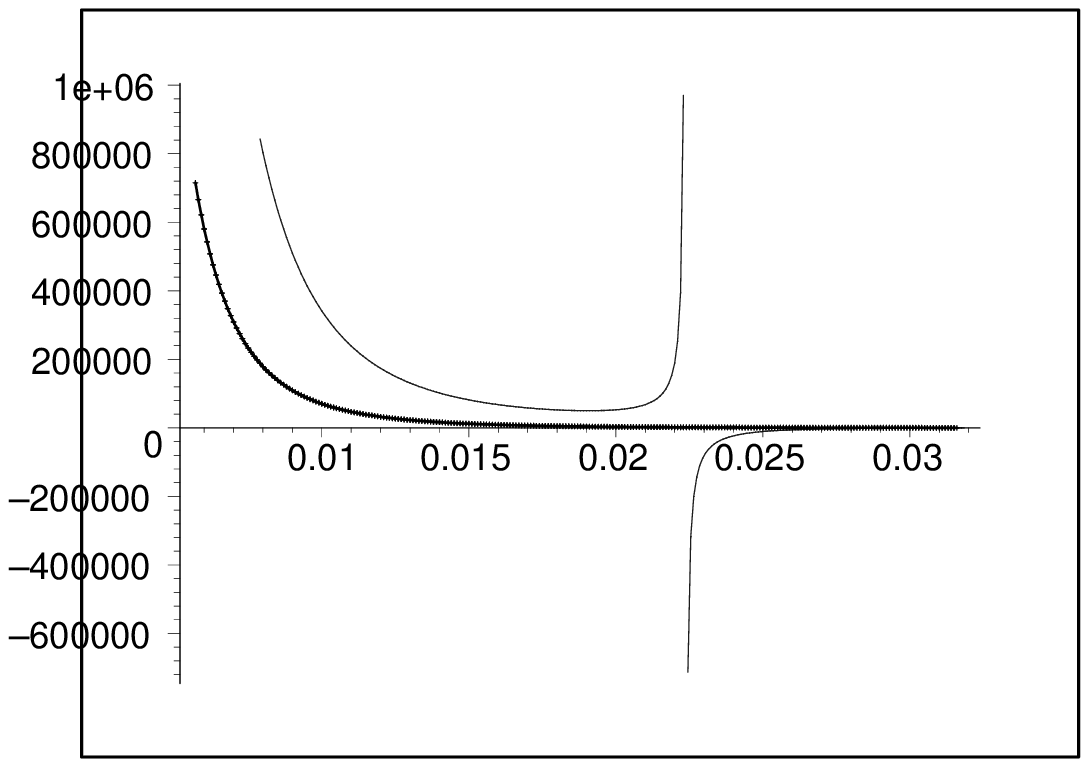}} \caption{Entropy
(bold line) and heat capacity (solid line) of 6-dimensional bolt
solution for $l=1$ and $V=1$ in the allowed range of $N$.}
\label{CS6b}
\end{figure}
\begin{figure}[tbp]
\epsfxsize=6cm \centerline{\epsffile{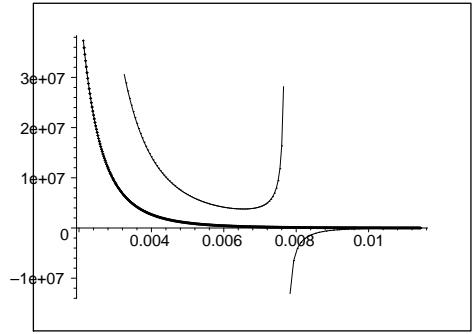}} \caption{Entropy
(bold line) and heat capacity (solid line) of 4-dimensional bolt
solution for $l=1$ and $V=3$ in the allowed range of $N$.}
\label{CS6c}
\end{figure}
Figures \ref{CS6a}, \ref{CS6b} and \ref{CS6c} show the entropy and
specific heat as a function of $N$ for $l=1$, $V=0$, $V=1$ and $V=3$ up to $%
N_{\max }$ for outer horizons of the bolt solutions respectively. As one can
see from these figures, the bolt solution is stable in the range $0<N<$ $N_{%
\mathrm{st}}<N_{\max }$, while $N_{\mathrm{st}}$ is less than $N_{\max }$
for $V\neq 0$, and equal to $N_{\max }$ for $V=0$. That is, the uncharged
solution is stable in the whole allowed range $0<N<$ $N_{\max }$, while the
stable phase becomes smaller as $V$ increases.

\section{Thermodynamics of $(2k+2)$-dimensional Charged Taub-NUT Solutions%
\label{2k+2-therm}}

Using Eqs. (\ref{Actg}), (\ref{Ict}) and (\ref{Iem}), the
Euclidean actions can be calculated as
\begin{equation}
I_{N_k}=-J(k)[l^2-2kN^2+2(-1)^{k+1}(2k-1)^2l^2V^2],
\end{equation}
where $J(k)$ is a function of $k$ as
\begin{equation}
J(k)=\frac{(4\pi )^{k+1}(k+1)\Gamma (\frac 12-k)\Gamma (k+1)}{16\pi ^{3/2}l^2%
}N^{2k}.
\end{equation}
Also the total mass may be calculated as
\begin{equation}
\mathcal{M}_n=\frac{k(4\pi )^{k-1}}2m_n.
\end{equation}
As in the case of uncharged solutions \cite{CFM}, the total angular momentum
is zero. The entropy can also be obtained through the use of Gibbs-Duhem
relation (\ref{GibD}), where $\mathcal{C}_i=Q$ and $\Gamma _i=\Phi $ as
\begin{equation}
S_n=-J(k)[(2k-1)l^2-2k(2k+1)N^2+2(2k-1)^3l^2V^2] \label{Sk}
\end{equation}
The Smarr-type formula for mass versus extensive quantities $S$
and $Q$ may be written as
\begin{equation}
\mathcal{M}_n=k(4\pi )^{k-1}B(k)Z^{(k-1/2)}[l^2-2(k+1)Z]+\frac{2kD(k)}{(4\pi
)^kZ^{(k-1/2)}}[\frac{\Gamma (k-\frac 12)}{\Gamma (k+1)(2k-1)}]^2Q^2,
\end{equation}
where $Z$\ is the solution of the following equation:
\begin{eqnarray}
&&S_n+\frac{(4\pi )^{k+1}(k+1)\Gamma (\frac 12-k)\Gamma (k+1)}{16\pi
^{3/2}l^2}\{(2k-1)l^2-2k(2k+1)Z  \notag \\
&&+\frac{(2k-1)l^2}{2\pi }[\frac{\Gamma (k-\frac 12)}{(4\pi )^{k-1}\Gamma
(k+1)Z^{(k-1/2)}}]^2Q^2\}Z^k=0.
\end{eqnarray}
One may then regard the parameters $S$ and $Q$ as a set of
extensive parameters for the mass $\mathcal{M}(S,Q)$ and define
the intensive parameters conjugate to $S$ and $Q$, which are the
temperature and the electric potential respectively. One obtains
\begin{eqnarray}
T_n &=&\left( \frac{\partial \mathcal{M}_n}{\partial S}\right) _Q=\frac
1{4(k+1)\pi N}, \nonumber\\
\ \ \Phi _n &=&\left( \frac{\partial M}{\partial Q}\right) _S=V_n.
\label{TPd}
\end{eqnarray}
Equations (\ref{TPd}) show that the the quantities $T_n$ and
$\Phi_n$ coincide with those which was calculated in Sec.
(\ref{d-dim}). Thus the thermodynamic quantities calculated in
Sec. (\ref{d-dim}) for Nut solution, satisfy the first law of
thermodynamics. The specific heat capacity may be obtained as
\begin{equation}
C_n=2J(k)[k(2k-1)l^2-2k(k+1)(2k+1)N^2-2(k-1)(2k-1)^3l^2V^2].  \label{C-k}
\end{equation}
Since $J(k)$ will produce a minus sign for odd $k$, Eqs. (\ref{Sk}) and (\ref
{C-k}) show that the entropy and heat capacity are positive for odd $k$
provided
\begin{equation*}
\frac{2k(2k+1)}{(2k-1)}N^2-2(2k-1)^2l^2V^2<l^2<\frac{%
2k(k+1)(2k+1)N^2+2(k-1)(2k-1)^3l^2V^2}{k(2k-1)}
\end{equation*}
Hence, for odd $k>1$, the NUT solutions will be thermally stable in the
above range, which becomes increasingly narrow with increasing
dimensionality and wide with increasing $V$. For $k=1$, as we see in Sec.
\ref{four-Dim}, there exists another stable phase. For even $k$, however, no
minus sign is produced by $J(k)$, meaning that the entropy will be positive
for
\begin{equation*}
l^2<\frac{2k(2k+1)}{(2k-1)}N^2-2(2k-1)^2l^2V^2
\end{equation*}
and the specific heat will be positive for
\begin{equation*}
l^2>\frac{2k(k+1)(2k+1)N^2+2(k-1)(2k-1)^3l^2V^2}{k(2k-1)}
\end{equation*}
Since the second value will always be larger than the first, this means
there is no range in which both will be positive, and thus for all even $k$,
the NUT solutions will be thermally unstable.

\section{Conclusion}

In this paper, we first, constructed asymptotically AdS Taub-NUT/bolt
solutions in ($2k+2$)-dimensional Einstein-Maxwell gravity with curved base
spaces. We found that the function $F(r)$ of the metric does not depend on
the specific form of the base factors on which one constructs the circle
fibration. In the presence of electromagnetic field, there exist two extra
parameters, in addition to the mass and the NUT charge, namely; the electric
charge $q$ and the potential at infinity $V$. We found that in order to have
NUT charged black holes in Einstein-Maxwell gravity, in addition to the two
conditions of uncharged NUT solutions, there exists two other conditions.
The first extra condition comes from the regularity of vector potential at $%
r=N$ which gives a relation between $q$ and $V$. Indeed, the existence of
the parameter $V$ enables us to get a regularity condition on the one-form
potential which is identical to that required to obtain a NUT solution. If
one of these parameters vanishes then the other one should be equal to zero
and the solution reduces to the uncharged solution. The second extra
condition comes from the fact that the horizon at $r=N$ should be the outer
horizon of the black hole. Indeed, Einstein-Maxwell gravity admits NUT black
holes provided the potential parameter is less than a critical value $V_{%
\mathrm{crit}}$, which may be obtained by solving the system of
two equations (\ref{gnut}). In any dimension larger than four, the
mass parameter $m$ which is fixed by these four NUT conditions
depends on the fundamental constant $l$ and parameters $N$ and $V$
(or $q$), while in 4-dimensions the mass parameter does not depend
on $V$ (or $q$). We also found that the NUT solutions have no
curvature singularity at $r=N$ when the metric of the base space
is chosen to be $\mathbb{CP}^{k}$, and have curvature singularity
for other curved base spaces. Then, we obtained the bolt solutions
of Einstein-Maxwell gravity in various dimensions, and gave the
equations which can be solved for the horizon radius of the bolt
solution. We found that in order to have bolt solution, the NUT\
parameter should be smaller than a maximum value $N_{\mathrm{\max
}}$ which decreases as $V$ increases.

Second, we investigated the thermodynamics of NUT solutions in any
dimensions. We calculated the conserved quantities and the Euclidean actions
of the NUT/bolt solutions through the use of counterterms renormalization
procedure. Also we obtained the charge and electric potentials of the
solutions in an arbitrary dimension, and calculated the entropy through the
use of Gibbs-Duhem relation. We obtained a Smarr-type formula for the mass
as a function of the extensive parameters $S$ and $Q$, calculated the
temperature and electric potential, and showed that these quantities satisfy
the first law of thermodynamics. Then, we studied the phase behavior of the
charged NUT solutions in canonical ensemble by calculating the heat
capacity, and found that the NUT solutions are not thermally stable for even
$k$, while there exists a stable phase for odd $k$, which becomes
increasingly narrow with increasing dimensionality and wide with increasing $%
V$. We also studied the phase behavior of bolt solutions in 4 and
6 dimensions in canonical ensemble and found that these solutions
have a stable phase, which becomes smaller as $V$ increases.\\

\begin{center}
\textbf{ACKNOWLEDGMENTS}
\end{center}
This work has been supported in part by Research Institute for
Astrophysics and Astronomy of Maragha, Iran.

\end{document}